# The combined design of climate mitigation and geoengineering

Wang Liang[1*], Huang Qiu-An[2]

[1] Department of Control Science and Engineering, Huazhong University of Science and Technology, WuHan, Hubei, 430074, P.R. China

[2] Faculty of Physics and Electronic Technology, Hubei University, Wuhan, Hubei, 430062, P.R. China

*To whom correspondence should be addressed. E-mail:wangliang.f@gmail.com

**[Abstract]** Combined climate mitigation/geoengineering approach has better economic utility, less emission control rate and temperature increase than mitigation alone. If setting the 50% reduction rate and 2 ℃ temperature increase as constrains, we find there is no a feasible solution for emission control, but combined design is still available.

Now there are mainly two kinds of methods to avoid the impacts of climate change. The first is mitigation method like reducing the CO2 output. The second is geoengineering method. These measures mainly include putting up space shields that cover billions of square meters, using chemicals to reflect sunlight or to increase Earth's cloud cover, or stimulating massive growth of phytoplankton in the oceans (1).

The mitigation method has been approved in Tokyo protocols and recent Copenhagen accord. And the geoengineering is still some theory research. But in these years, geoengineering begin to attract more and more attentions for two reasons: first, the efficacy of mitigation may not enough to counteract the extremely climate crisis (2). Second, with the improvement of technology, the cost of geoengineering may become affordable.

Mitigation and geoengineering are studied in different areas. Only few researches concern their combinations (3). For example, Wigley shows the combined mitigation and geoengineering design could 'buying time' for a policy needed to stabilize concentrations through a simple climate model (4). We study the combined design by integrated assessment model (IAM), which couples sub models of the climate and economic systems. Using this kind of model, we could determine an 'optimal' policy by balancing the benefits and cost of mitigation and geoengineering methods.

We chose Nordhaus' Dynamic Integrated Climate and Economy (DICE) model for our analysis because of its relative simplicity and transparency (5). The mitigation methods are all carried out by reducing carbon emission.

Assuming the geoengineering method is space shields. Its benefit is featured in F(t), the increased radiative forcing caused by greenhouse gas in DICE. Let Geo(t) denote the space shield control. It could shield at most 1.8% income solar radiation, about 3.7w/m2 radiative forcing. So the new 'radiative forcing':

$$F^*(t) = F(t) - Geo(t), 0 \leq Geo(t) \leq 3.7$$

Its cost is subtracted from C(t), the consumption in DICE. For 1.8% shield, the cost per year averages to 100 billion dollars (6). The period in DICE is 10 years. So the geoengineering cost of per w/m2 is about 0.2703 trillion in one period. The new consumption:

$$C^*(t) = C(t) - 0.2703 Geo(t)$$

We set all the parameters as the default values of DICE2007 and run the model from 2005 to 2205. Then we compare the combined design with emission control design. We find the economic utility of combined design (152057) is more than emission control design (151237). The combined design also has less emission control requirement and temperature increase (Fig.1).

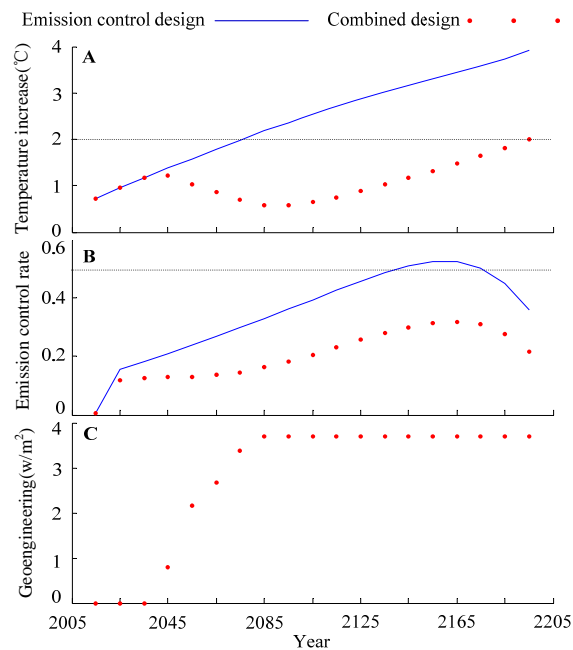

Fig.1 (A) the temperature increase, (B) emission control rate, (C) geoengineering control. The dot is the combined design and the line is the emission control design.

Then we consider more actual condition. First, it's difficult to reach the international climate agreement with more than 50% reduction rate in carbon dioxide emission. Second, the 2℃ temperature increase is regarded as the safe level. Exceeding it may result in some catastrophic climate changes. Considering these two constrains, we design the climate policy again.

We find there is no a feasible solution for model with only emission control, mainly because the maximal control rate can't ensure all the temperature increases bellow the safe level. But the combined design is still available. It is very similar to the combined design in Fig.1.

Our results suggest that combined design of geoengineering and mitigation may be a promising strategy for counteracting the climate change. But now geoengineering is still regarded as an unsafe option (7). For example, it risks major precipitation changes. Its impact to ecosystem is also unclear. So we may consider the geoengineering only if warming causes sufficiently harmful impacts.

**References and Note**